\documentstyle[prl,aps,epsfig,floats]{revtex}
\begin{document}
\draft
\newcommand{\beq}{\begin{equation}}
\newcommand{\eeq}{\end{equation}}
\newcommand{\bea}{\begin{eqnarray}}
\newcommand{\eea}{\end{eqnarray}}
%%%%%%%%%%%%%%%%%%%%%%%%%%%%%%%%%%%%%%%%%%%%%%%%%%%%%%%%%%%%%%%%%%%%%%%
\def\npb#1#2#3{Nucl. Phys. B{#1} (19#2) #3}
\def\plb#1#2#3{Phys. Lett. B{#1} (19#2) #3}
\def\PLBold#1#2#3{Phys. Lett. {#1B} (19#2) #3}
\def\prd#1#2#3{Phys. Rev. D{#1} (19#2) #3}
\def\prl#1#2#3{Phys. Rev. Lett. {#1} (19#2) #3}
\def\prt#1#2#3{Phys. Rep. {#1} C (19#2) #3}
\def\moda#1#2#3{Mod. Phys. Lett.  {#1} (19#2) #3}
%%%%%%%%%%%%%%%%%%%%%%%%%%%%%%%%%%%%%%%%%%%%%%%%%%%%%%%%%%%%%%%%%%%%%%%
\def\lsim{\raise0.3ex\hbox{$\;<$\kern-0.75em\raise-1.1ex\hbox{$\sim\;$}}}
\def\gsim{\raise0.3ex\hbox{$\;>$\kern-0.75em\raise-1.1ex\hbox{$\sim\;$}}}
\def\Frac#1#2{\frac{\displaystyle{#1}}{\displaystyle{#2}}}
\def\al{\alpha}
\def\be{\beta}
\def\ga{\gamma}
\def\de{\delta}
\def\si{\sigma}
\def\C{{\cal{C}}}
\def\O{{\cal{O}}}
\def\wt{\widetilde}
\def\ol{\overline}
\def\l{\left}
\def\r{\right}
\def\no{\nonumber\\}
%%%%%%%%%%%%%

\twocolumn[\hsize\textwidth\columnwidth\hsize\csname@twocolumnfalse\endcsname

\title{Flavour-Dependent CP Violation and Natural Suppression of the Electric Dipole Moments}
\author{S. Abel$^1$, D. Bailin$^1$, S. Khalil$^{1,2}$, and
O. Lebedev$^1$\\}

\address{$^1$ Centre for Theoretical Physics, University of Sussex, Brighton BN1
9QJ,~~~U.~K.} 
\address{$^2$ Faculty of Science, Ain Shams University, Cairo, 11566, Egypt.}
\date{\today}  
\maketitle
\begin{abstract} 
We revisit the supersymmetric CP problem and find that it can be naturally resolved
if the origin of CP violation is closely related to the origin of flavour structures.
In this case, the supersymmetry breaking dynamics do not bring in any {\em new}
CP-violating phases. This mechanism requires hermitian Yukawa matrices which naturally
arise in models with a U(3) flavour symmetry. 
The neutron electric dipole moment (NEDM)
is predicted to be within one-two orders of magnitude below the current experimental limit.
The model also predicts
a strong correlation between $A_{CP}(b\to s \gamma)$ and the NEDM.
The strong CP problem is mitigated although not completely solved.
\end{abstract} 
\pacs{PACS numbers:11.30.P, 11.30.E, 13.40.E
\hfill SUSX-TH-00-022
}
\vskip.5pc
]
%%%%%%%%%%%%%%%%%%%%%%%%%%%%%%%%%%%%%%%%%%%%%%%%%%%%%%

The tight experimental bounds on the EDMs are an important tool for
understanding supersymmetry breaking. For example, they might indicate
that the SUSY parameters happen to fall in small regions of the
parameter space where there are accidental cancellations between
various contributions or where the supersymmetric phases are
small \cite{cancellations}.  Another  possibility is that the sfermions
of the first two generations have masses in the TeV
range. Alternatively, and perhaps more naturally, these bounds might
indicate that the {\em flavour-independent} SUSY phases are forced to
vanish by the underlying theory. If this is correct, it seems likely
that CP violation is intimately related to the {\em origin} of the
flavour structures in the model rather than the origin of
supersymmetry breaking.  In other words the source of CP violation is
associated with whatever generates the SUSY Yukawa interactions, and
the supersymmetry breaking dynamics do not contribute any further CP
violation. Under this assumption it may be  possible to redefine the 
fields such that the moduli and dilaton  auxiliary fields  have {\em real}
vacuum expectation values (VEVs).
Such  behaviour has been observed in some explicit string models in
which CP violation is derived from first principles in effective type
I string models~\cite{SAGS}.

Since the CP breaking scale is related to the scale at which the
flavor symmetry gets broken, it must be far above the SUSY breaking
scale, probably close to the string or GUT scales.  If the underlying
structure indeed allows only quantities with a non-trivial flavor
structure to have CP-violation, this automatically implies that the
gaugino masses and the $\mu$ term are {\em real} (we illustrate it below).
On the other hand,
the trilinear soft couplings may still involve CP-violating phases,
and ${\cal{O}}(1)$ phases of the diagonal elements of the $A$--terms
induce unacceptably large EDMs by themselves. Following the above
logic, we conclude that they too are prohibited by the mechanism
generating CP violation.  An important point to note however, is that
this is generally a {\em basis dependent} requirement, which is to say
that the diagonal $A$--terms must be real in the quark mass basis.
Conversely, even if the $A$--terms are all real to begin with,
diagonalizing the quark Yukawas generally causes phases to appear in the
diagonal terms.

This fact strongly suggests that the flavour structure  forces the
diagonal $A$--terms to be real in {\em any} basis.  In what follows we
consider the only possibility of which we are aware, that the
flavour structure is hermitian.  Hermitian Yukawa matrices  naturally occur
in models with a horizontal flavour symmetry \cite{masiero} and  left-right symmetric models
\cite{mohapatra}.
  As we shall shortly see,
if SUSY and CP breaking are decoupled, this guarantees a nearly
hermitian structure for the $A$--terms and hence suppresssed EDMs.
This setting  appears in string models with hidden sector
fields in the {\em adjoint} representation of certain symmetry groups. 

Models with non-universal A-terms have recently 
attracted considerable attention \cite{nonuniversal}.
In these models, in order to avoid a conflict with the EDM
bounds, the condition of small diagonal phases 
(or EDM cancellations) is normally imposed
by hand without referring to a particular mechanism. In
contrast, the EDM in models with hermitian flavour structures
is a derived quantity. We
find that the neutron EDM is predicted to be within one or two orders of
magnitude below the current experimental limit.  Also, we show that
the SUSY CP-phases in this class of models can have a significant
effect on the B and K physics observables, in particular the
$b\rightarrow s \gamma$ CP-asymmetry and the $K^0-\bar K^0$ mixing.
An interesting feature of this scenario is a strong correlation between
$A_{CP}(b\to s \gamma)$ and the  neutron EDM, i.e. a large ($\sim 10\% $) CP-asymmetry
in  the  $b\rightarrow s \gamma$ decay implies that the NEDM is of order 
$10^{-26}e\cdot cm$ which is just below the current limit.
This prediction can be tested in the near future.

We will study the implications of
hermitian Yukawa matrices in the supergravity framework:
\begin{eqnarray}
&& Y^{u \dagger}=Y^u \;,\;  Y^{d \dagger}=Y^d \;,\;  Y^{l \dagger}=Y^l \;,\; \nonumber\\
&& {\rm Arg}(\mu)={\rm Arg}(M_i)=0 \;.
\end{eqnarray}
In supergravity models, the soft SUSY breaking parameters are given in
 terms of the K\"ahler potential and the superpotential. In
 particular, the trilinear parameters are written as \cite{brignole}
\begin{eqnarray}
&& A_{\alpha \beta \gamma} = F^m \bigl[ \hat K_m  + \partial_m \log Y_{\alpha \beta \gamma}
-\partial_m \log (\tilde K_{\alpha} \tilde K_{\beta} \tilde K_{\gamma} ) \bigr] \;.
\label{A-terms}
\end{eqnarray}
Here the Latin indices refer to the hidden sector fields while the
Greek indices refer to the observable fields; the K\"ahler potential
is expanded in observable fields as $K=\hat K + \tilde K_{\alpha}
\vert C^{\alpha} \vert^2 +...$ and $\hat K_m \equiv
\partial_m \hat K$.		

Note that $\tilde K_{\alpha}$ is always real; $\hat K_m$ is real if
$\hat K$ is a function of $h_m+h_m^*$. The F-terms are real according
to the basic assumption of the model that the SUSY breaking dynamics
do not violate CP. For definiteness, let us  fix the index $\alpha$ to
refer to the Higgs fields, and indices $\beta$ and $\gamma$ to refer to
the left-handed and right-handed fields, respectively. The resulting $A$--terms 
are hermitian in  the generational indices 
if the derivatives of the K\"ahler potential $\tilde K_{\beta , \gamma}$
  are either generation-independent (for the left
and right fields separately) or the same for the left and right fields of the
same generation ($F^m {\hat K_m}+  F^m \partial_m \log
Y_{\alpha \beta \gamma}$ is hermitian). 
These conditions are satisfied in simple Type I and heterotic string models.
Consequently, the quantities $\hat
A_{\alpha \beta \gamma} \equiv A_{\alpha \beta \gamma} Y_{\alpha \beta
\gamma}$ are hermitian with respect to the generational indices; i.e.
\begin{eqnarray}
&& \hat A^{u \dagger}=\hat A^u \;,\;  \hat A^{d \dagger}=\hat A^d \;,\;  \hat A^{l \dagger}=\hat A^l
 \;.\;
\end{eqnarray}
The flavour structure of the $\hat A$'s can be quite different from
that of the $Y$'s. Note that even if the Yukawa couplings have a
negligible modular dependence and $A_{\alpha \beta \gamma}$ are real
(and symmetric), the quantities $\hat A$ responsible for the
phenomenology must involve complex phases.  The other soft breaking
parameters are real as they are independent of $Y_{\alpha \beta
\gamma}$. In what follows, we will assume for simplicity a universal soft scalar
mass parameter\footnote{This is the case if the only non-trivial flavour structure
in the model comes from the Yukawa matrices or when the derivatives of the K\"ahler potential
are small compared to $m_{3/2}^2$.}.

All these quantities are subject to  renormalization group (RG) 
running which may not preserve
the hermiticity.  For instance, the RG evolution of $Y^u$ involves a
term $Y^d Y^{d \dagger} Y^u$ which is not hermitian; the amount of
generated ``non-hermiticity'' depends on $[Y^u,Y^d]$ and therefore is
suppressed by the off-diagonal entries of the Yukawas. Generally, the
Yukawa running effects are negligible for the first two generations
anyway, so potentially non-negligible non-hermitian contribution can
come only from the third generation. Similar considerations apply to
the RG evolution of the $A$--terms. These radiative effects induce
non-zero EDMs as discussed below.  The phases of the gaugino masses
and the $\mu$-term are RG-invariant and remain zero.

Therefore, the Yukawas and A-terms stay hermitian to a good degree
even at low energies.  The hermitian Yukawas are diagonalized by a
unitary transformation,
\begin{eqnarray}
&& u \rightarrow V^u u\;,\; Y^u \rightarrow V^{u\dagger} Y^u V^u
\end{eqnarray} 
and similarly for the down quark and lepton fields. The resulting CKM
matrix is given by $V_{CKM}=V^{u\dagger}V^d$. If we transform the
sfermion fields in the same way as we tranform the fermion fields, we
will go over to a basis known as the ``super-CKM'' (SCKM) basis in the
literature. The $A$--terms transform accordingly;
\begin{eqnarray}
&&\hat A^u \rightarrow V^{u\dagger} \hat A^u V^u \;,\; \hat A^d \rightarrow V^{d\dagger} \hat A^d V^d
\;.
\label{transform}
\end{eqnarray} 
The $A$--terms in this super-CKM basis remain hermitian and therefore
have real diagonal elements.  As a result, the flavour-conserving mass
insertions appearing in the EDM calculations
 $(\delta_{ii}^{d(u)})_{LR}= {1\over \tilde m^2}
 ( (\hat A^{d(u) \dagger}_{SCKM})_{ii} v_{1(2)} - Y_i^{d(u)} \mu
v_{2(1)})$ (no sum over $i$) are real to a good degree. This provides
a natural suppression of the EDMs.  Note that hermiticity of the
Yukawas is crucial for this suppression. If we give it up, $A$--terms
which are hermitian or even real at high energies will gain diagonal
phases due to the transformation of the type (\ref{transform}) with
$V_L \not = V_R$.

%-----------------------
\begin{figure}
\epsfig{file=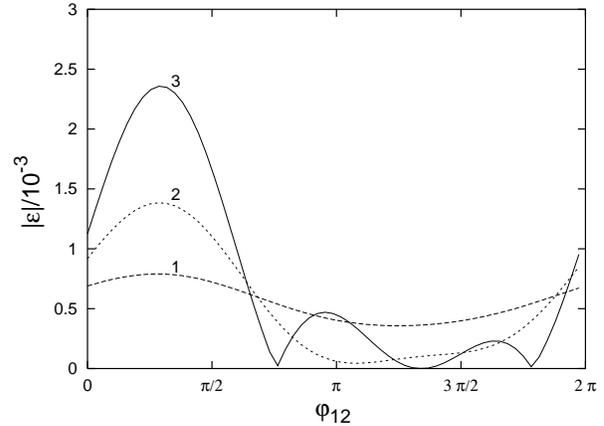,width=8cm}\\

\caption{SUSY contribution to $\vert \varepsilon \vert $ as a
function of the  phase $\phi_{12}$  for $m_0 \simeq 150$ GeV,
$\tan \beta =3$, and $m_{\chi^+}\simeq 100$ GeV. Curve 1: $\vert A_{12}\vert=
m_0$; curve 2: $\vert A_{12}\vert=3m_0$; curve 3: $\vert A_{12}\vert=5m_0$.
The observed value is $\vert  \varepsilon \vert \simeq 2.26 \times 10^{-3} $.  }
\label{fig1}
\end{figure}
%-----------------------

Our numerical studies show that the radiative effects producing
non-hermitian pieces in the Yukawas and $A$--terms typically lead to a
neutron EDM between $10^{-27}$ and $10^{-28}e\cdot cm$ and an electron EDM of
about $10^{-33}e\cdot cm$. In some cases, when the condition of a
large CP-asymmetry in  $b\rightarrow s \gamma$ is imposed, the NEDM can be
as large as ${\cal{O}}(10^{-26})e\cdot cm$. 
Generically the model predicts the NEDM
to be within two orders of magnitude below the current experimental bound.

It is interesting to note that the strong CP problem is mitigated in
this model. The $\bar \theta$ parameter vanishes both at the tree and
the 1-loop leading log levels. This can be seen from the fact that
${d\over dt} {\rm Det}Y={\rm Det}Y \;{\rm Tr} \left[ Y^{-1} {d\over
dt} Y \right]$ with ${\rm Tr} \left[ Y^{-1} {d\over dt} Y \right]$
being real (see the RGEs in \cite{Bertolini:1991if}).  The gluino phase is
RG-invariant and vanishes at the high energy scale.  As a result, the
$\bar \theta$ parameter remains zero under the RG flow.  However,
there are one loop finite contributions which generate both a gluino
phase and ${\rm Arg(Det}Y^{u,d})$. The strong CP problem in our model
is milder than it normally is in SUSY models, yet it still exists.

The complex phases appearing in the off-diagonal elements of the
$A$--terms can have a significant impact on the kaon and B physics.
In our numerical studies we used the following representative
GUT scale hermitian  Yukawa matrices

\begin{equation}
\small{Y^u=\left( \matrix {4.1\times 10^{-4}&6.9\times 10^{-4}\;{\rm i}&-1.4\times 10^{-2} \cr
-6.9\times 10^{-4}\;{\rm i}&3.5\times 10^{-3}&-1.4\times 10^{-5}\;{\rm i}\cr
-1.4\times 10^{-2}&1.4\times 10^{-5}\;{\rm i}&6.9\times 10^{-1}
             } \right)\; },
\end{equation}

\begin{eqnarray}
\small{Y^d=\!\!\left(\!\! \matrix {1.3\!\times\! 10^{-4}\!\!\!&\!\!\!(2.0 \!
+\!1.8\; {\rm i}) \!\times\! 10^{-4}&\!\!-4.4\!\times\! 10^{-4} \cr
(2.0\!-\!1.8\;{\rm i})\!\times\! 10^{-4}&9.3\!\times\! 10^{-4}&
7.0\!\times\! 10^{-4}\;{\rm i}\cr
-4.4\!\times\! 10^{-4} & -7.0\!\times\! 10^{-4}\;{\rm i} & 1.9\!\times\! 
10^{-2}}\!\right),}
\nonumber
\end{eqnarray}
%\begin{eqnarray}
%&& Y^u=\left( \matrix {4.1\times 10^{-4}&6.9\times 10^{-4}\;{\rm i}&-1.4\times 10^{-2} \cr
%-6.9\times 10^{-4}\;{\rm i}&3.5\times 10^{-3}&-1.4\times 10^{-5}\;{\rm i}\cr
%-1.4\times 10^{-2}&1.4\times 10^{-5}\;{\rm i}&6.9\times 10^{-1}
%             } \right)\; , \nonumber \\
%&& Y^d=\left( \matrix {1.3\times 10^{-4}&2.0\times 10^{-4}+1.8\times 10^{-4}\;{\rm i}&-4.4\times 10^{-4} \cr
%2.0\times 10^{-4}-1.8\times 10^{-4}\;{\rm i}&9.3\times 10^{-4}&
%7.0\times 10^{-4}\;{\rm i}\cr
%-4.4\times 10^{-4} & -7.0\times 10^{-4}\;{\rm i} & 1.9\times 10^{-2}
%             } \right)\; ,
%\end{eqnarray}
which at low energies reproduce the quark masses and the CKM matrix (we use $\tan\beta=3$
throughout the paper). In this analysis it is necessary to use the complete set of the MSSM
RG equations as given in Ref.\cite{Bertolini:1991if}.

 We
find that the SUSY contribution to the $\varepsilon'$ parameter
is negligible. This occurs due to severe cancellations between the
contributions involving $(\delta_{12}^d)_{LR}$ and
$(\delta_{12}^d)_{RL}$ mass insertions (we use the standard
definitions of \cite{Gabbiani:1996hi}). Due to the hermiticity
$(\delta_{12}^d)_{LR} \simeq (\delta_{12}^d)_{RL}$, whereas they
contribute to $\varepsilon'$ with opposite signs.

On the other hand, the
SUSY contribution to the $\varepsilon $ parameter can be substantial.
The gluino--down squark contribution to  $\varepsilon $ does not suffer from similar
cancellations.
In Fig. 1 we plot the values of $\vert \varepsilon \vert $ versus the phase ($\phi_{12}$)
of the off--diagonal element $A_{12}$ for three representative values of 
$\vert A_{12} \vert$, namely $\vert A_{12} \vert = m_0$ (curve 1),  
$\vert A_{12} \vert = 3 m_0$ (curve 2), and  $\vert A_{12} \vert = 5 m_0$ (curve 3).
For simplicity we assume $A^d = A^u =A^l$. We fix the universal scalar mass $m_0$ to 
be  150 GeV and adjust $m_{1/2}$ such that the lightest chargino mass is around 
100 GeV.  We have also fixed the other elements of the A-terms to be $m_0$. 

The major contribution to $\varepsilon$ comes from the gluino--down squark diagrams with 
the left-right and right-left mass insertions which in this case interfere constructively.
 The left-left and right-right mass insertions
appear only due to the RG running effects. Their typical values are ${\cal{O}}(10^{-5})$ and
 ${\cal{O}}(10^{-6})$, respectively, which are  too small to produce any significant effect.
The value of  $(\delta_{12}^d)_{LR}$  which saturates the 
observed  $\vert  \varepsilon \vert \simeq 2.26 \times 10^{-3} $ is given by $\sqrt{\vert
\mathrm{Im}(\delta_{12}^d)^2_{LR} \vert } \simeq 3.5 \times 10^{-4}$
for the gluino and squark masses of  500 GeV \cite{Gabbiani:1996hi}.
Fig.1 demonstrates that increasing the GUT
value of $\vert A_{12} \vert $ increases 
the SUSY contribution to $\varepsilon $, as expected.
 Note that  curve 1 never crosses the horizontal axis
 meaning that
the SUSY contribution to $\varepsilon $ in this case does not vanish for any value of
$\phi_{12}$. This occurs due to the non-trivial phases in the Yukawa matrices.
The main message of these results is that the  SUSY contribution can 
accommodate large values of $\varepsilon $ without violating the EDM constraints. 
Indeed, we
found that the NEDM is about $2\times 10^{-27}e\cdot cm$ in all of these cases. As we discuss
below, these results have implications for the B system as they affect the prediction
for the $B^0-\bar B^0$ mixing phase.

In the B system, the observables of primary
interest are the CP-asymmetry in the $b\rightarrow s \gamma$ decay and the
$B^0-\bar B^0$ mixing. 
In Fig.2 we present a plot showing the dependence of
$A_{CP}(b\rightarrow s \gamma)$ on  the phase $\phi_{23}$ of the element $A_{23}$ for 
three values of $\vert A_{23} \vert=\vert A_{33} \vert$:
 $\vert A_{23} \vert = m_0$ (curve 1),  
$\vert A_{23} \vert = 3 m_0$ (curve 2), and  $\vert A_{23} \vert = 5 m_0$ (curve 3).
Other elements are taken to be $m_0$.
We automatically impose the condition of the correct branching ratio, so the 
points not satisfying this constraint are not shown.
For values of $\vert A_{23} \vert $ close to $m_0$, the CP-asymmetry is  a few percent 
which is already significantly larger than the SM expectation. 
As emphasised in Ref.\cite{bailin}, the CP--asymmetry 
is very sensitive to the magnitude of the elements $A_{23}$ and $A_{33}$.
For a sufficiently large $\vert A_{23} \vert=\vert A_{33} \vert$ ($\geq 3 m_0$),
the CP-asymmetry can be as large as 6-8 percent.  It could be even larger in other
regions of the parameter space.

The magnitudes of  $A_{23}$ and $A_{33}$ have a significant impact on the RG evolution 
of the A-terms. In particular, they are responsible for generating complex phases 
in $(\delta_{11}^{d(u)})_{LR}$
which produce the NEDM. We found that curve 3 corresponds to the NEDM of $4.1\times 10^{-26}e\cdot cm$,
while for curves 2 and 1 it is $2.6\times 10^{-26}e\cdot cm$ and $2.5\times 10^{-27}e\cdot cm$, 
respectively. This provides an interesting
signature of the model: a large CP-asymmetry in the $b\rightarrow s \gamma$ decay can be 
accommodated only if the NEDM is just below the current limit. 
 
The SUSY contribution to the  $B^0-\bar B^0$ mixing is small for typical values of the A-terms
(i.e. $\vert (\delta_{13}^d)_{LR}\vert \lsim 5\times 10^{-4}$).
Nevertheless, supersymmetry can affect it indirectly by changing the predicted value for
$\sin 2 \beta$ \cite{silva}.
 As we have shown, SUSY may be responsible for the 
observed value of $\varepsilon$. 
If it is, there is no constraint from $\varepsilon$ on the CKM matrix.
In this case
lower values for $\sin 2 \beta$ are allowed \cite{mele}, in better agreement with the
recent BaBar and Belle results \cite{babar}.

%-----------------------
\begin{figure}
\epsfig{file=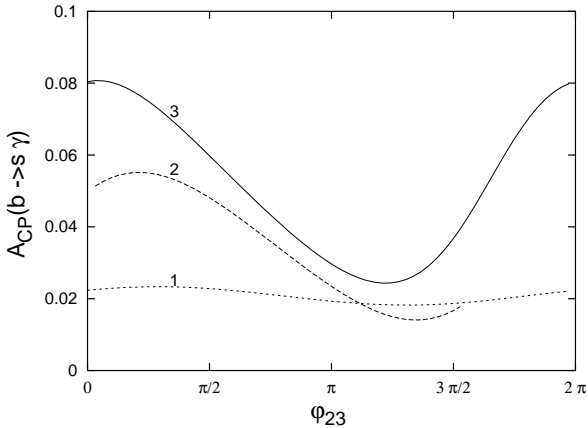,width=8cm}\\

\caption{CP-asymmetry $A_{CP}(b\to s \gamma)$ as a
function of the  phase $\phi_{23}$  for $m_0 \simeq 150$ GeV,
$\tan \beta =3$, $m_{\chi^+}\simeq 100$ GeV, and $\vert A_{23}\vert=\vert A_{33}\vert$.
 Curve 1: $\vert A_{23}\vert=
m_0$; curve 2: $\vert A_{23}\vert=3m_0$; curve 3: $\vert A_{23}\vert=5m_0$.}
\label{fig2}
\end{figure}
%-----------------------

Now we turn to the discussion of how hermitian Yukawa matrices can be
implemented.  One of the possibilities is based on a concept of
so-called ``real CP-violation'' introduced by Masiero and Yanagida
\cite{masiero}, and Lavoura \cite{lavoura}.  Hermitian Yukawa matrices
may appear due to a (gauged) horizontal symmetry $U(3)_H$ which gets
broken spontaneously by the VEVs of the 
real adjoint fields $T^a_{\alpha}$
($a=1..9;\; \alpha=u,d,l,..)$.  Some of these {\em real} VEV's also break
CP since some of the components of $T^a_{\alpha}$ are CP-odd.  As a
result, CP violation appears in the superpotential through complex
Yukawa couplings, whereas the $\mu$-term is real since it arises from a $U(3)_H$
invariant combination of the $T$'s. An effective  $U(3)_H$-invariant superpotential of the type
$ \hat W = {g_H\over M}  \hat H_1 \hat Q_i ( T^a_d\lambda^a)_{ij} \hat D_j$
produces the Yukawa matrix 
$(Y^d)_{ij}= {g_H\over M} \langle T^a_d \rangle (\lambda^a)_{ij}$, where $\lambda^0$
is proportional to the unit matrix  and $\lambda^{1-8}$ are the Gell-Mann matrices.

In a somewhat different context, hermitian Yukawa matrices and real flavour-independent quantities 
arise in the left-right symmetric models \cite{mohapatra}. The left-right symmetry may
be too restrictive to satisfy all of the phenomenological constraints
even if one allows for a significant fine-tuning
(see the last reference in \cite{mohapatra}). The B-factories' data will be necessary to
make a conclusive statement concerning the viability of such models.

In view of the current high level of activity in the subject,
 it is of interest to see whether
 and how the fields needed may arise in string theory. 
Real adjoint fields may only arise from non-supersymmetric states, while complex
adjoints could be part of a chiral multiplet (in which case hermitian Yukawas 
appear only  if all the VEVs are real).
Massless, adjoint, chiral
multiplets occur in (weakly coupled) heterotic
string models when the affine lie algebra is realized at level
$k\geq 2$. (See, for example, \cite{FIQ90}.) Such models enable the 
construction of  ($k=2$) string GUTs based on $SU(5)$ or $SO(10)$, although 
realistic models are difficult to obtain \cite{AFIU95},\cite{AFIU96}. However,
in none of the models is there more than one such adjoint multiplet, and their 
construction is such that the scalars arise as continuous moduli, having no self coupling,
and whose VEVs must therefore be fixed by non-perturbative effects. Adjoint
matter representations  also arise in Type IIB orientifold models with discrete torsion 
\cite{KR173}, both as a chiral field in a supersymmetric sector (of the $Z_2 \times Z_4$ model with 
$U(8)$ gauge group), and as a scalar field in a non-supersymmetric sector (of the  $Z_2 \times Z_6$
model with  $USp(8) \times U(4)$ gauge group). Either of these can, in principle, accommodate 
the standard model gauge group together with a $U(3)$ horizontal symmetry, but as before in 
neither is there more than one copy of the adjoint. 
Currently, we are not aware of string models in which all of the required features arise.
This subject requires further study.

This work was supported by PPARC.


\begin{thebibliography}{99}

\bibitem{cancellations} J. Ellis, S. Ferrara, D.V. Nanopoulos, Phys. Lett. B {\bf 114}, 
231 (1982); W. Buchm\"{u}ller and D. Wyler, Phys. Lett. B {\bf 121}, 321 (1983);
J. Polchinski and M.B. Wise, Phys. Lett. B {\bf 125}, 393 (1983);
T. Ibrahim and P. Nath, Phys. Lett. B {\bf 418}, 98 (1998); 
Phys. Rev. D {\bf 57}, 478 (1998); Phys. Rev. D {\bf 58}, 111301 (1998);
T. Falk and K.A. Olive, Phys. Lett. B {\bf 439}, 71 (1998);
M. Brhlik, G.J. Good and G.L. Kane, Phys. Rev. D {\bf 59}, 115004 (1999);
M. Brhlik, L. Everett, G.L. Kane and J. Lykken, Phys. Rev. Lett. {\bf 83}, 2124 (1999);
A. Bartl, T. Gajdosik, W. Porod, P. Stockinger and H. Stremnitzer,
Phys. Rev. D {\bf 60}, 073003 (1999); S. Pokorski, J. Rosiek and C.A. Savoy,
Nucl. Phys. B {\bf 570}, 81 (2000).

\bibitem{SAGS} S.~A.~Abel and G.~Servant,
hep-th/0009089; and {\em in preparation}.

\bibitem{masiero}
A.~Masiero and T.~Yanagida, hep-ph/9812225.

\bibitem{mohapatra}
R.~N.~Mohapatra and G.~Senjanovic,
Phys.\ Lett.\ {\bf B79}, 283 (1978);
R.~N.~Mohapatra and A.~Rasin,
Phys.\ Rev.\ Lett.\ {\bf 76}, 3490 (1996);
Phys.\ Rev.\ {\bf D 54}, 5835 (1996);
K.~S.~Babu, B.~Dutta and R.~N.~Mohapatra,
Phys.\ Rev.\ {\bf D 61}, 091701 (2000).


\bibitem{nonuniversal}
S.A. Abel and J.-M. Fr\'{e}re, Phys. Rev. D {\bf 55}, 1623 (1997);
S. Khalil, T. Kobayashi and A. Masiero, Phys. Rev. D {\bf 60}, 075003 (1999);
S. Khalil and T. Kobayashi, Phys. Lett. B {\bf 460}, 341 (1999);
S. Khalil, T. Kobayashi and O. Vives, Nucl. Phys. B {\bf 580}, 275 (2000);
M. Brhlik, L. Everett, G.L. Kane, S.F. King, O. Lebedev,
Phys. Rev. Lett. {\bf 84}, 3041 (2000). 

\bibitem{brignole} A. Brignole, L. Ib\'{a}\~{n}ez, C. Mu\~{n}oz, in
{\it Perspectives on Supersymmetry}, edited by G.L. Kane (World Scientific,
Singapore,1998), hep-ph/9707209. 

\bibitem{Bertolini:1991if}
S.~Bertolini, F.~Borzumati, A.~Masiero and G.~Ridolfi,
Nucl.\ Phys.\  {\bf B353}, (1991) 591.

\bibitem{Gabbiani:1996hi}
F.~Gabbiani, E.~Gabrielli, A.~Masiero and L.~Silvestrini,
Nucl.\ Phys.\  {\bf B477}, 321 (1996).

\bibitem{bailin} D. Bailin and S. Khalil, hep-ph/0010058.

\bibitem{silva} J. Silva and L. Wolfenstein, hep-ph/0008004;
see also A. Masiero, M. Piai, O. Vives, hep-ph/0012096.

\bibitem{mele} S. Mele, Phys. Rev. D {\bf 59}, 113011 (1999).

\bibitem{babar} D. Hitlin, BaBar collaboration, talk given at ICHEP'2000 
(Osaka, Japan), SLAC-PUB-8540; H. Aihara, Belle collaboration, {\it ibid.}

\bibitem{lavoura}
L. Lavoura, Phys. Lett. B {\bf 400}, 152 (1997);
Phys. Rev. D {\bf 61}, 076003 (2000).

\bibitem{FIQ90} 
A.~Font, L.~E.~Ib$\rm {\acute{a}}\rm{\tilde{n}}$ez and F.~Quevedo,
Nucl.~Phys. {\bf B345}, 389 (1990).

\bibitem{AFIU95} 
G.~Aldazabal, A.~Font, L.E.~Ib$\rm {\acute{a}}\rm{\tilde{n}}$ez 
, and A.M.~Uranga, Nucl. Phys. {\bf B452}, 3 (1995).

\bibitem{AFIU96} 
G.~Aldazabal, A.~Font, L.E.~Ib$\rm {\acute{a}}\rm{\tilde{n}}$ez 
, and A.M.~Uranga, Nucl. Phys. {\bf B465}, 34 (1996).

\bibitem{KR173} 
M.~Klein and R.~Rabadan, JHEP {\bf 0010}, 049 (2000).


\end{thebibliography}
\end{document}